\documentstyle[12pt,epsf]{article}

\def\be{\begin{equation}}
\def\bea{\begin{eqnarray}}
\def\ee{\end{equation}}
\def\eea{\end{eqnarray}}
\def\ri{\rightarrow}

\def\ra{\rangle}
\def\la{\langle}
\def\r{\right}
\def\l{\left}

\def\a{\alpha}
\def\t{\tau}
\def\g{\gamma}
\def\G{\Gamma}
\def\s{\sigma}

\begin{document}
\title{Attractors in fully asymmetric neural networks}
\author{U. Bastolla$^{1,}$ $^2$ and G. Parisi$^1$}
\maketitle
\centerline{$^1$Dipartimento di Fisica, Universit\`a ``La Sapienza'', P.le Aldo
Moro 2, I-00185 Roma Italy}
\centerline{$^2$HLRZ, Forschungszentrum J\"ulich, D-52425 J\"ulich Germany}
\medskip
\centerline{Keywords: Disordered Systems, Attractor Neural Networks}

\begin{abstract}
The statistical properties of the length of the cycles and of the weights of
the attraction basins  
in fully asymmetric neural networks ({\it i.e.} with completely uncorrelated
synapses) are computed in the framework of the annealed approximation which we
previously introduced for the study of Kauffman networks. Our results show 
that this model behaves essentially as a Random Map possessing a reversal
symmetry. Comparison with numerical results suggests that the approximation
could become exact in the large size limit.
\end{abstract}

\section{Introduction}
In the past decade Attractor Neural Networks were the subject of an intense
study as a model of associative memory. The "ancestor" of these models,
the Hopfield model \cite{Hop}, was defined as follows: there is a set of $N$
neurons, each one associated with a binary variable $\s_i\in\{0,1\},
i\in\Omega=\{1,\cdots N\}$, representing its activity.
The synaptic couplings between these
model neurons, $J_{ij}$, are chosen at random at the beginning and kept
fixed, and then the system evolves deterministically according to the equation
\be \s_i(t+1)={\rm sign}\l(\sum_j J_{ij}\s_j(t)\r) \label{evol}\ee
(parallel updating; alternatively, one can consider sequential updating when
$\s_i(t+1)$ is determined by the state of $\s_j(t+1)$ for $j<i$ and by
$\s_j(t)$ for $j>i$).

This procedure defines a disordered dynamical system: the evolution is
deterministic, but its rules are chosen at random at the beginning and
kept fixed. In other words, we can rewrite the dynamic law in the form

\be {\cal C}(t+1)=f_J\l({\cal C}(t)\r), \label{dis}\ee
where ${\cal C}$ represents a configuration of the system, {\it i.e.}
a set of values of the $N$ variables $\s_i$, $f_J$ is a random
realization of a deterministic map and the set of indices $J$ labels
the realization of the dynamic rules. 

The most natural distance in configuration space is the
normalized Hamming distance, defined as
\be d(C,C')={1\over N}\sum_i \mid \s_i-\s_i'\mid.\ee

We are interested in the statistical properties of the motion
asymptotically in time and system size. As the motion is deterministic and
configuration space is finite, asymptotically in time the dynamics
takes place on periodic orbits, and the quantities of interest are the
lengths and the number of such orbits as well as the size of their
attraction basins. Such quantities are random variables, depending on
the realization of the dynamical rules, and we will study their
probability distribution.

When the couplings are symmetric ($J_{ij}=J_{ji}$) it is possible to define a
Hamiltonian so that equation (\ref{evol})
represents the zero temperature dynamics of a thermodynamic system. In
particular, if the $J_{ij}$ are chosen from a distribution with zero mean
and variance $1/N$ (for instance a Gaussian distribution) we are dealing with
the zero temperature dynamics of the SK model (in the case of sequential
updating: the parallel updating does not imply a relaxational dynamics).
In the Hopfield model the couplings are symmetric, too, but they are chosen
according to the Hebbian rule:

\be J_{ij}=\sum_{\mu=1}^P \xi_i^\mu \xi_j^\mu, \ee
where the $P$ vectors of $N$ binary variables, ${\bf \xi}^\mu$, represent the
memorized patterns. The system is able to memorize, in the sense that the
patterns are fixed points of the dynamics and they are stable against random
perturbations if their number does not exceed the capacity of the network,
{\it i.e.} if $P$ is not larger than $\a_c N$, with $\a_c\approx 0.14$.
So, given a number of microscopic states growing as $2^N$, the Hopfield
model is able to memorize a number of patterns growing linearly with
$N$.

Asymmetric neural networks received a large attention in the literature in
the late '80s \cite{CS1,CS2,GRY,Hei1,Hei2,Hei3}. In 1986 it was  proposed to
generalize the Hopfield model by taking into
account also asymmetric couplings \cite{P86}.This generalization
appears more realistic, since synapses in nature are in general not
symmetric, and it suggests a possible way to distinguish  between a
network that has remembered a learned pattern and a network which is
in a confused state (such a distinction is not possible in the
Hopfield model). In fact, in asymmetric neural networks, two kind of
attractors are present: ``ordered" attractors, that are either short
cycles or fixed points, and ``chaotic" attractors, whose length grows
exponentially with system size. The first numerical observations of
this twofold nature of the attractors are due to Gutfreund, Reger and
Young \cite{GRY} and to N\"utzel \cite{Nu}.

In this note we are mainly interested in the study of the
properties of the attractors, such as the probability distributions of
their lengths, of their number and of the size of their attraction basins.
We will consider only the case of fully asymmetric couplings, {\it i.e.}
$J_{ij}$ and $J_{ji}$ are independent random variables.
In this case analytical results have already been obtained about the
correlation functions \cite{Hei1,Hei2} and about the number of
attractors \cite{eta0}, but more about the attractors can be said
using a simple stochastic scheme based on the annealed
approximation. This approximation was introduced in the study of disordered
dynamical systems by Derrida and Pomeau \cite{DP} to study damage
spreading in Kauffman networks (a disordered dynamical system proposed
as a model of the genetic regulation in cells \cite{K69}). In
\cite{BP0} we showed that it can also be used to obtain information
about the attractors of that model.

A reason of interest of this study is that asymmetric neural
networks are the limit case of a one parameter family of models, the
parameter $\eta$ representing the symmetry of the synaptic couplings:

\be  \eta={\l\la J_{ij}J_{ji}\r\ra\over \l\la J_{ij}\r\ra^2}.\ee

The case $\eta=0$ represents the present model  (fully asymmetric couplings),
while for $\eta=1$ the couplings are fully symmetric and we
obtain the mean field model of spin glasses. Thus the parameter $\eta$
connects with continuity asymmetric neural networks to a disordered
system of statistical mechanics.

It was suggested through numerical simulations that the model with
generic correlation undergoes a dynamical transition when $\eta$ is
changed \cite{Nu,NK,CFV}. The transition seems to take place when the
absolute value of $\eta$ crosses the value $1/2$. For $|\eta |< 1/2$
the dynamics is chaotic and the typical length of the cycles increases
exponentially with the number of neurons $N$, while for $|\eta|>1/2$
the dynamics is frozen and the typical length of the cycles does not increase
with system size (most of the cycles have length 2, for positive
$\eta$, and 4 for negative $\eta$). 

This transition is reminiscent of the dynamical transition
taking place in Kauffman networks. Also in that case the typical
length of the cycles  grows exponentially with $N$ in the so called
chaotic phase, remains finite in the frozen phase and grows less than
exponentially with $N$ on the critical line \cite{K69,BP0}.
It was claimed by Kauffman that the critical line of his model can be a good
model of the genetic regulatory systems acting in cell differentiation, thus
showing that such systems do not need to be tuned in the very details by
natural selection but behave similarly to typical realizations of an ensemble
of random regulatory networks \cite{K69}. It is possible that,
analogously, also the supposed critical point in attractor neural
networks, where chaotic and ordered cycles coexist, can suggest
something interesting from a biological point of view. We
think that our method can be modified to give information
about systems with generic asymmetry and about the supposed phase
transition that they undergo, though this probably requires to go beyond the
annealed approximation.


\section{Closing probabilities}
\subsection{General framework}
Our strategy for the study of attractors in disordered dynamical
systems has as starting point the
probability distribution of the distance at different time
steps. Actually, the information contained in the distribution of
the distance is much more than what we need and this distribution is
in principle a very complicated object, so that our approach may seem
to complicate the problem. But, in some cases, the distance can be
well approximated by a suitably defined stochastic process and
the computation becomes much easier. The simplest possibility is to
approximate the distance with a Markovian stochastic process. This is
what we call here the {\it annealed approximation}.

An apparently paradoxical aspect of this approach is that in
disordered dynamical systems attractors exist due to the fact that the
motion is deterministic. Stochastic processes, on the other hand, have
nothing similar to a limit cycle. Nevertheless, all the properties of
the attractors can be derived from the distribution of distances,
which is a well defined object in both kinds of models. We can not
pursue this analogy up to times larger than the time of first
recurrence of a configuration already visited, when the deterministic
motion becomes periodic. But this is enough, since the first
recurrence provides us with every information about the length of the cycles
and the transient time.

The fundamental object of our study will be then the distribution of
distances between configurations at time steps $t$ and $t'>t$ on the
same trajectory, restricted to trajectories that have not
yet visited twice any configuration up to the larger time $t'$ (thus
the effects of periodicity do not yet appear). We
will call this condition the opening condition, and denote it by the
symbol $A_{t'}$.  The closing probability $\pi_N (t,t')$ is the
probability that configurations at time steps $t$ and $t'$ are equal
($d(t,t')=0$), subject to the opening condition:

\be \pi_N (t,t')=\Pr\l\{ d(t,t')=0 \mid A_{t'}\r\} \ee
(the subscript $N$ is there to remember the dependence on system size).

After the closing time $t'$ the trajectory enters a periodic orbit of length
$l=t'-t$, where $t$ is the transient time. In terms of the closing
probabilities, the probability to find such a trajectory is easily
computed. First we have to know the probability $F_N(t)$ that the
trajectory was not closed before time $t'=t+l$. This obeys the equation
$F_N(t+1)=F_N(t)\l(1-\sum_{t'=0}^{t-1}\pi_N(t',t)\r)$, whence,
introducing a continuous time variable, we get 

\be F_N(t)=\exp\l(-\int_0^t dt'\int_0^{t'} dt^{''}
\pi_N(t',t^{''})\r)\label{FN}\ee 
(to have a slightly simpler formula we made the hypothesis that
the typical closing times are long, which is normally the case in the chaotic
phase, where they grow exponentially with system size $N$, and we
transformed the sum into an integral).
 
The probability to find a trajectory that, after a transient time $t$,
enters a cycle of length $l$ is then obtained multiplying $F_N(t+l)$ times
the closing probability $\pi_N(t,t+l)$. 

\subsection{The annealed approximation}\label{ann}
Regarding the distance as a Markovian stochastic process is a very
drastic approximation. In our case it sounds reasonable when the
temporal distance $l$ is large, since the
model that we study is known to have a behavior very reminiscent of
chaos \cite{CFV}. However in this way we neglect some memory
effects, which can play a fundamental role in systems with nonzero
symmetry. 

This approximation was first used in this
context by Derrida and Pomeau \cite{DP}, who studied the damage
spreading in Kauffman networks. They showed that the average value of
the Hamming distance between two different trajectories is
equivalent, in the infinite system limit, to the average value
of the Markovian stochastic process obtained extracting new dynamical
rules at every time step and keeping memory only of the value of the
distance at time step $t-1$. Thus the disorder is treated as annealed
rather than as quenched. In other words, instead of considering an
ensemble of trajectories, each one taking place on a fixed realization
of the dynamical rules, they consider an ensemble of trajectories
moving from one realization of the dynamical rules to another one, in
the same spirit in which the annealed average is used for disordered
thermodynamical systems. 


The above procedure can be shown to describe exactly the evolution of
the average distance up to time of order $\log N$ in disordered
systems with finite connectivity \cite{DW,HN}, but we think that its validity
is more general. In systems with infinite connectivity like the one that we
are studying here, or when one is interested in the whole distribution of the
distance, the equivalence between the two dynamics has not been proved, and we
have to assume that a typical trajectory of the quenched system loses
memory of the details of the realization of dynamical rules under which it
evolves. In the Random Map model \cite{RM} this is trivially true. 
In other cases this can be thought of as a maximal ignorance hypothesis,
whose consequences must then be compared with numerical simulations.

Let us state some of these consequences. A Markovian stochastic
process, if its transition probability is ergodic\footnote{In the
  present case, in order to have an ergodic 
transition probability, we must exclude as starting point the distance
$d=0$ which is an absorbing point (if $d(t,t+l)=0$, we must have with
probability one $d(t',t'+l)=0$ for every $t'\ge t$), that is we have
to impose the condition that the trajectory is not yet closed, as we did.},
converges to a stationary stochastic variable independent of the
initial distribution. This means that the closing probability
$\pi_N(t,t+l)$ converges to a stationary value $\pi_N^*$. This is also
independent of $l$ if the transition probability does not depend on
this quantity. We will show that this happens in the present case, at
least for $l$ large enough.

It is then easy to compute the probability of a trajectory which,
after a transient time $t$, enters a cycle of length $l$ (with $l$ and
$t$ large enough, so that the closing probability has reached its
asymptotic value): using the results of last section, we get

\be \Pr\l\{T=t,L=l\r\}= {1\over\t^2_N}\exp\l(-{1\over 2}\l({t+l\over
  \t_N}\r)^2\r),  \label{Pcic}\ee
where $\t_N={\pi^*_N}^{-1/2}$ is the typical time scale of the problem, in the
sense that the random variable $t/\t_N$ has a well defined density of
probability even in the limit where $\t$ goes to infinity.
All the dependence on system size is contained into the factor $\pi^*_N$,
which is expected to decrease exponentially with $N$ in the chaotic phase.
For instance, for a uniform Random Map \cite{RM}, which is the most chaotic
disordered dynamical system, it holds $\pi^*_N=1/2^N$, and
consequently the typical time scale of the attractors grows as
$2^{N/2}$.

The properties of Random Maps can be easily generalized starting from
equation (\ref{Pcic}). 
An interesting quantity is the distribution of the attraction basin
weights. This  was analytically computed by Derrida and Flyvbjerg for
the case of the uniform RM \cite{DF}. The weight of the attraction
basin of cycle $\a$, $W_\a$, is defined as the
probability to extract at random an initial configuration which will
reach asymptotically the attractor $\a$. The statistical information
about the distribution of the weights can be expressed through the
``moments" $\la Y_n\ra$, defined as

\be \l\la Y_n\r\ra =\sum_\a \l\la W_\a^n\r\ra. \ee

$Y_1$ is equal to 1 due to the normalization of the
weights and $Y_2$ represents the average weight in a given
dynamical system. Its extreme values, 1 and 0, correspond respectively
to the ``ergodic" case where there is only one relevant attractor and to the
case where there is an infinite number of relevant attractors, while a
finite value of $Y_2$ means that there is a finite number of
attractors with non vanishing weight. This quantity fluctuates from
sample to sample, so it is necessary to consider an average over the
realizations of the dynamical rules, that is represented by the
angular brackets.

The method used in \cite{DF} to compute the distribution of the weight
can be applied without modifications to
all disordered dynamical systems where the closing probability reaches
an asymptotic value, $\pi^*_N$, and the result do not depend on this
value in the large size limit. Thus the distribution of the attraction
basin weights is {\it universal} for all the disordered dynamical systems
where the closing probability reaches a stationary value
\cite{BP0,BP1}, apart for systems which possess some
symmetry. The result for the average value of the $Y_n$ is \cite{DF}

\be \la Y_n\ra={4^{n-1}\l[(n-1)!\r]^2\over (2n-1)!}. \label{Yn}\ee

The fluctuations from sample to sample can also be computed. For
example the fluctuations of $Y_2$ are measured by $\la Y^2_2\ra -\la
Y_2\ra ^2$, and do not cancel even in the infinite size limit $N\ri\infty$.

\vspace{.5cm}
The average number of attractors of length $l$ can be computed
starting with the relation

\be  \l\la n_a(l)\r\ra={2^N\over l}\Pr\{T=0,L=l\}.\label{nl}\ee

In this formula, the probability $\Pr\{L=l,T=0\}$ should be computed
multiplying $F_N(l)$, given in equation (\ref{FN}),
times the closing probability $\pi_N(0,l)$. This is different from
the asymptotic value (for large $t$) of $\pi_N(t,t+l)$. According to
the hypothesis, on which the annealed approximation relies, that the
system is going to lose memory of the details of the evolution, we
expect no correlations between the initial configuration and a
configuration at a large time $l$ or, in other words, we expect the
closing probability to be, asymptotically in $l$, $\pi_N(0,l)=1/2^N$.
For chaotic Kauffman networks this can be explicitly computed in the
framework of the annealed approximation, which is then consistent
under this point of view. Moreover, we find in this case
$\pi_N(0,l)=c_l 1/2^N$, where $c_l$ does not depend on $N$. Thus it holds
\be \l\la n_a(l)\r\ra\approx {c_l\over l}\exp\l(-l^2/2\t^2\r).\ee

Summing over $l$ we obtain the average value of the total number of cycles,
whose leading term in $N$ is equal to $\log \t_N$.
Since in the chaotic phase the time scale grows exponentially with
system size, the number of attractors is proportional to $N$ in
this case. This computation holds for chaotic Kauffman networks in the
framework of the annealed approximation, but we expect it to hold more
generally under the hypothesis discussed above.

\subsection{Master equation}\label{dist}
The general scheme described above must be modified in the case studied in
this note, to take into account the symmetry of the problem. Let us define
the reversal operator, $R$, which reverses all the spins. This
operator commutes with the dynamics of the system. Using the notation
defined in equation (\ref{dis}), we can write:

\be f_J \l( R{\cal C}\r)=R f_J\l({\cal C}\r), \label{sym}\ee

This implies that we can define two different  closing times:

\begin{enumerate}
\item The first time $t$ when ${\cal C}(t)$ is equal to ${\cal C}(t+l)$.

\item The first time when ${\cal C}(t+l)$ is equal to $R{\cal
  C}(t)$: then equation (\ref{sym}) implies ${\cal C}(t+2l)={\cal C}(t)$. In
  other words, the trajectory has reached, after a transient time $t$, a
  cycle of length $2l$.
\end{enumerate}

These closing events can be described in terms of the Hamming distance
between configurations: the first one corresponds to  $d(t,t+l)=0$, while the
second one corresponds to $d(t,t+l)=1$. Thus two closing probabilities
must be defined:

\bea \pi_N^{(0)}(t,t')=\Pr\l\{d(t,t')=0\mid A_{t'}\r\}, \\
\pi_N^{(1)}(t,t')=\Pr\l\{d(t,t')=1 \mid A_{t'}\r\},
\eea
and the opening condition, $A_t$, has the meaning that up to time $t$
it never occurred either $d(t_1,t_2)=0$ or $d(t_1,t_2)=1$.
Our task is now to compute the master equation for the distribution of the
distance under the opening condition (this means that we consider
only trajectories not yet closed) and under the hypothesis
that the distribution of $d(t+1,t'+1)$ depends only on the distribution of
$d(t,t')$. This is not a difficult task. To simplify slightly the
formulas we will consider, instead of the distance, the overlap
$q=1-d$, which is measured by the number of elements whose state is the same
in the two configurations, divided by $N$.

An element $\s_i$ is in the same state at time $t+1$ and $t'+1$ if its local
field  has the same sign at time steps $t$ and $t'$. Thus it holds

\be\s_i(t+1)\s_i(t'+1)={\rm sign}\l(\sum_{jk}J_{ij}J_{ik}\s_k(t)\s_j(t')\r).
\ee

Let us consider separately the contribution to this sum coming
from the spins whose state is the same at time steps $t$ and $t'$,
whose number is $Nq(t,t')$, and that belong to a set that we indicate with
the name $I(t,t')$. We can then write

\be \s_i(t+1)\s_i(t'+1)={\rm sign}\l((h_i^+(t))^2-(h_i^-(t))^2\r), \ee
where
\bea  && h_i^+(t)=\sum_{j\in I(t,t')}J_{ij}\s_j(t), \label{h+-}\\ 
&& h_i^-(t)=\sum_{j\in\Omega /I(t,t'))} J_{ij}\s_j(t). \nonumber\eea 
 
The annealed approximation consists in considering the local fields as random
variables, correlated to the previous story of the system only through
the value of $q(t,t')$. In
this spirit, we consider a dynamics in which the local fields are
extracted at random at every time step, under the following assumptions:

\begin{enumerate}
\item The local fields at different points are independent random variables;
\item The value of $\s_j(t)$ is independent on the synaptic coupling 
$J_{ij}$. 
\end{enumerate}

Both these assumptions have troubles when the synaptic couplings are correlated
with each other, but they are quite reasonable for $\eta=0$,
which is the case that we are studying now.
Assumption 1 implies that the transition probability is a binomial one:

\be \Pr\l\{q(t+1,t'+1)=q_n\mid q(t,t')=q_m\r\}={N\choose n}\l(\g(q_m)\r)^n
\l(1-\g(q_m)\r)^{N-n}, \ee
where $q_n=n/N$, and $\g(q)$ is the probability that 
$|\tilde h^+(q)|>|\tilde h^-(1-q)|$, where, following assumption 2, 
$\tilde h^\pm (q)$ is a Gaussian variable with mean value zero and variance
$q$ (this result is independent on the details of the distribution of the
couplings, provided that they are all independent variables with mean value
zero and with the same variance). A straightforward computation shows that

\be \g(q)={2\over \pi}\arcsin \sqrt q. \ee

The Markov process associated to this transition probability is
ergodic if we exclude as starting points the values $q=0$ and $q=1$,
as we do imposing the opening condition, and the distribution of the
distance evolves towards a stationary distribution. Moreover, since
the transition probability is independent on $l=t'-t$, also the
stationary distribution is independent on $l$, which appears only in
the initial distribution of the variable $q(0,l)$.
It is also evident from the symmetry of the problem that it must hold $\g(q)= 
1-\g(1-q)$, so, if also the initial distribution is symmetric ({\it
  e.g.} a binomial distribution around $q=1/2$), the overlap
distribution will be symmetric at every time step and it will be
concentrated  around the value $Q(t,t')=D(t,t')=1/2$ (the distributions of 
the overlap and of the distance are perfectly equivalent in this
case). The stationary distribution is, independently on the initial
one, concentrated around the value $Q^*$ solution of the
self-consistent equation:

\be Q^*=\g\l(Q^*\r)={2\over \pi}\arcsin \sqrt{Q^*}.\label{fix}\ee
This equation has three solutions: $1/2$, 1 and 0, but only the first
one can be accepted, according to the criterion $|\g'(Q^*)| < 1$,
which can be obtained either as the stability condition of the
fixed point $Q^*$ of the map $Q(t+1,t'+1)=\g(Q(t,t'))$, or as the
condition that the variance of the stationary distribution is positive (see
equation (\ref{Var}) below).

Equation (\ref{fix}) is equivalent to the equation for the
stationary value of the correlation function rigorously derived in
\cite{Hei1} through a functional integral approach, so that one can
see from this comparison that the annealed approximation gives an
exact (though trivial) result concerning the average overlap. But our
task here is to compute the whole stationary distribution of
the overlap, and we can not prove that the annealed approximation is
correct to this extent, so we have to rely upon simulations to control
its validity.

Though it is concentrated around $Q^*=1/2$, the stationary
distribution is much broader than a binomial one and thus the closing
probability is exponentially larger than $1/2^N$. In order to compute
its value, we proceed in this way \cite{BP0}: since the
transition probability is exponentially concentrated, we look for a
solution of the form:

\be P_N\l(q(t,t')=q_n\r)=C_N(q_n,t)\exp\l(-N\a_t(q_n)\r), \ee
where we have dropped the $l$ dependence of the probability, which disappears
at stationarity. Using Stirling approximation for the binomial coefficient
and the saddle point approximation to average over the distribution at time
step $t-1$, we get the following equation for the evolution of the exponent
of the distribution, $\a_t(x)$:
\be \a_t(x)=\a_{t-1}(q_t(x))+x\log\l({x\over \g(q_t(x))}\r)+
(1-x)\log\l({1-x\over 1-\g(q_t(x))}\r), \label{alp} \ee
where the function $q_t(x)$ must be determined self consistently solving the
equation
\be \a'_{t-1}(q_t(x))-\g'(q_t(x))\l({x\over \g(q_t(x))}-{1-x\over
1-\g(q_t(x))}\r), \label{sella}\ee
with the conditions $q_t(x)>0$ and $q_t(x)<1$.

At stationarity the most probable overlap (the point where $\a_t(q)$ has
a minimum) is given by equation (\ref{fix}), and the variance of the
distribution can be obtained taking the second derivative of equation
(\ref{alp}) and solving it together with the first derivative of
the saddle point condition (\ref{sella}). The result is
                  
\be V^*={Q^*\l(1-Q^*\r)\over 1-\l(\g'(Q^*)\r)^2}={1/4\over
  1-(2/\pi)^2}\approx 0.4204, \label{Var}\ee
where $V^*$ is the variance of the stationary distribution multiplied
times $N$. Thus the variance is larger than in the case of a
binomial distribution, since the dynamics has produced correlations
  between different elements. 

The value of the closing probability can not be computed analytically:
we need for this the whole function $\a(x)$, and to obtain it we
should solve a transcendent non local equation. Thus we had to solve
numerically equation (\ref{alp}), obtaining the stationary
distribution reported in Figure 1. The asymptotic closing probability,
defined as $P_N^*(q=1)+P_N^*(q=0)$, is thus
\be \pi^*_N=2 \exp\l(\a N\r), \ee
with $\a=0.4554$. As discussed in the previous section, the exponent
of the average length of the cycles should be equal to $\a/2$. This prediction
is in good agreement with the numerical simulations that will be
reported in section 4.

\subsection{Initial distribution}

In order to compute the average number of cycles we have to know the
distribution of the overlap with the initial configuration, $q(0,l)$,
which plays the role of the initial distribution for the stochastic
process studied in the previous subsection. Although the number of
cycles in fully asymmetric neural networks was already exactly studied
by Schreckenberg \cite{eta0}, we want to sketch the annealed
computation of it, since it is much simpler and it can be generalized
to more complex situations.

Our aim is to compute the distribution of $q(0,l)$. After one time
step the annealed approximation is exact (we have still to extract all the
couplings) and trivial: every spin can be either in its initial state
or in the reversed one with probability 1/2, and the overlap $q(0,1)$
multiplied times $N$ has a binomial distribution with $p=1/2$.
After two time steps we distinguish two contributions in the local
field: one coming from the set $I_1$ of the spins which are in the
same state at $t=0$ and at $t=1$ and another one coming from all the
other spins. We write

\be \s_i(0)\s_i(2)={\rm sign}\l(\sum_{j\in I_1}\s_i(0) J_{ij}\s_j(0)-
\sum_{j\in \Omega/I_1}\s_i(0) J_{ij}\s_j(0)\r). \label{Pl1}\ee

Since the states $\s_i(0)$ and $\s_j(0)$ are independent both one on
each other and on the couplings, we can set them equal to 1.
If we change the sign of the last sum, we obtain
$\s_i(0)\s_i(1)$. Thus, depending on whether this is positive or
negative, there are two possibilities:

\be \label{Pl2}
\s_i(0)\s_i(2)= \l\{
\begin{array}{lr}{\rm sign}\l(\l(\sum_{j\in I_1}J_{ij}\r)^2-
\l(\sum_{j\in \Omega/I_1}J_{ij}\r)^2\r), & if\; \s_i(0)\s_i(1)>0 \\
{\rm sign}\l(\l(\sum_{j\in \Omega/I_1}J_{ij}\r)^2-
\l(\sum_{j\in I_1}J_{ij}\r)^2\r), & if\; \s_i(0)\s_i(1)<0
\end{array}\r.\ee
(there is indeed in this formula a small imprecision, which becomes
negligible in the infinite size limit: since the coupling $J_{ii}$ is
set equal to zero, we have not to take into account the spin $j=i$
itself, which contributes to the first sum in both lines). The
probability that the sum of $n$ Gaussian variables has a module larger
than that of the sum of $N-n$ other Gaussian variables was already computed in
the previous section, where it received the name $\g(n/N)$.
Taking all this into account, we come to the transition probability

\bea & & \Pr\l\{q(0,2)=m/N\mid q(0,1)=n/N\r\}= \\
& & \hspace{2cm} =\sum_k {n\choose k} {N-n\choose m-k}
 \l(\g(n/N)\r)^{N-n-m+2k}\l(1-\g(n/N)\r)^{n+m-2k}, 
\nonumber\eea
where, as usual, the opening condition imposes to exclude as starting
points $n=0$ and $n=N$, and the sum runs over all the values of $k$
for which the factorial is well defined. The closing probability which
can be deduced from this formula setting either $m=N$ or $m=0$
coincides with the one exactly computed in ref. \cite{eta0}. It can be
easily seen that it is proportional to $1/2^N$, as expected (the
system loses memory of the initial configuration quite fast), and the
proportionality coefficient can be computed with the saddle point
method \cite{eta0}

In the general case, the information about $q(0,l)$ is not enough to
compute the distribution of $q(0,l+1)$: we have also to know the value
of $q(0,1)$, as it can be seen from equation (\ref{Pl2}) where we have
to substitute 1 with $l$ and 2 with $l+1$ in the equations but we have
to keep memory of $\s_i(0)\s_i(1)$ in the conditions. In the general case
the transition probability has thus the form

\bea & & \Pr\l\{q(0,l+1)=m/N\mid q(0,1)=n_1/N,q(0,l)=n/N\r\}=\\
& & \hspace{2cm}=\sum_k {n_1\choose k}{N-n_1\choose m-k}
\l(\g(n/N)\r)^{N-n_1-m+2k}\l(1-\g(n/N)\r)^{n_1+m-2k}, 
\nonumber\eea
and we have to consider the evolution of the joint distribution of the
variables $q(0,1)$ and $q(0,l)$. As expected, the correlations between
these two variables vanish very fast as $l$ grows, and the stationary
distribution is the product of two binomial distributions, as it can
be easily checked, so that for large $l$ the closing probability is
$\pi_N^{(a)}(0,l)=1/2^N$ (with $a$ equal either to 1 or to 0),
consistently with the supposed loss of
memory and in agreement with the exact results of ref \cite{eta0}. For
small values of $l$ it can be seen that $\pi_N^{(a)}(0,l)=c_l/2^N$,
where $c_l$ goes to a finite value in the infinite size limit, so that
the total number of cycles increases only proportionally to system size.

\section{Reversal symmetry}
The computations shown in section \ref{ann} must be modified
to take into account the twofold nature of the closing probability. We
have to distinguish between two kinds of cycles, with different
properties under the reversal operation:

\begin{enumerate}
\item Cycles that close when ${\cal C}(t+l)=R{\cal C}(t)$ (or, in
  other words, $q(t,t+l)=0$), whose length is $2l$. They are invariant
  under the reversal operation: each configuration is present together
  with its reversed one.

\item Cycles that close when  $q(t,t+l)=1$. In this case the reversal
  operator applied to the cycle $\G$ produces a new cycle $R \G$ with equal
  length and equally large attraction basin.
\end{enumerate}

Taking this into account, we have to distinguish between cycles of
even length, which can be of one of the two kinds, and cycles of odd
length, which can be only of the first kind. Cycle
length distribution is then

\bea \Pr\l\{T=t,L=l\r\} &&={1\over 2\t^2}\exp\l(-{(t+l)^2\over 2\t^2}\r),
\; l\:\: {\rm  odd}; \\
&&={1\over 2\t^2}\exp\l(-{(t+l)^2\over 2\t^2}\r)+{1\over 2\t^2}
\exp\l(-{(t+l/2)^2\over2\t^2}\r),\; l\:\: {\rm even}, \eea
with $\t=1/\sqrt\pi^*=1/\sqrt 2\exp(0.2277 N)$.

The cycles of the first type have only even length, so that their number
is half of the number of the cycles of the second type. Using the
result of subsection \ref{ann} and summing up the contributions of
both types of cycles we obtain, at the leading order in $N$,

\be \sum_l\la n_a(l)\ra\approx {3\over 2}\log\t={3\over 4}\a N, \ee
which is $3/2$ times larger than in a Random Map with the same closing
probability.

The most important difference between attractors in Asymmetric Neural
Networks and in a Random Map involves the distribution of the attraction
basins weights. Let us consider separately cycles of the first type
and cycles of the second type (taking only one cycle to represent each
pair of cycles of the second type). We then get the expression of the moments
$\la Y_n\ra$ of the distribution of the weights:

\be \la Y_n\ra={1\over 2}\l\la\sum_{\a'} W_{\a'}^n +
2\sum_{\a^{''}}\l(W_{\a^{''}}/2\r)^n\r\ra, \ee
where the sum over $\a'$ and $\a^{''}$ of the weights are both
normalized to one. Under the hypothesis that each of the two sets of
weights is distributed as in a Random Map, we get
\be \la Y_n\ra=\l({1\over 2}+{1\over 2^n }\r)\la Y_n\ra_{RM},
\label{weigh}\ee
or, using (\ref{Yn}),
\be \l\la Y_{n+1}\r\ra={1\over 2} {\l(n!\r)^2\over \l(2n+1\r)!}
\l(4^n+2^n\r). \ee

Thus the moments of the distribution of the weights are smaller than in the
usual Random Map, for instance $\la Y_2\ra=1/2$ instead of 2/3. These
results are in very good agreement with numerical simulations.

To prove equation (\ref{weigh}) let us recall that
$\la Y_n\ra$ can
be interpreted as the probability that $n$ randomly chosen
trajectories reach the same attractor. We can compute such quantity
using the closing probabilities and following exactly the same lines
as in \cite{DF}, but we have to remember that a closing event has two
different meanings: either a closure on an identical configuration
($q=1$) or a closure on a reversed configuration ($q=0$). Thus not all
the events which represent the closure of the $n$ trajectories, and
whose probability is exactly $\la Y_n\ra_{RM}$, have the meaning that
the trajectories will ultimately meet. If the first trajectory closes
with $q=0$ (this happens with probability 1/2), its attraction basin
contains also all of the reversed configurations, and the following $n-1$
trajectories which close on it will then go to the same attractor,
regardless on how they close. On the contrary, if the first
trajectory closes with $q=1$, the following $n-1$ trajectories have
also to close with $q=1$ in order to go to the same attraction basin
(if they close with $q=0$, they go to the reversed basin). In this
case, whose probability is again $1/2$, a closing event is equivalent to an
asymptotic meeting of the $n$ trajectories only with probability
$1/2^{n-1}$. Equation (\ref{weigh}) is thus proved. 


\subsection{Explicit symmetry breaking}
The above picture of the distribution of the attraction basins weights is
completely destroyed by the introduction of a magnetic field, however small,
in the equations of motion (\ref{evol}), which restores the
distribution typical of a Random Map. We  considered the dynamic rules
\be \s_i(t+1)={\rm sign}\l(\sum_j J_{ij}\s_j(t)-h\r) \label{evolh}\ee

The magnetic field $h$ has the biological meaning of the threshold of
activation of the neurons. In real neurons, such non-zero threshold exists
and can be different from one neuron to another one. In our simplified model,
we take a threshold which is constant among the different neurons. Its
introduction explicitly breaks the symmetry respect to the reversal of all
the neural activities.

In the framework of the annealed approximation, the conditional probability
that the activity of a neuron is the same in two different time steps is not
more symmetric, {\it i.e.} $\g(1-q)$ is different from $1-\g(q)$ and
$\g(1)$ increases very fast respect to the above case, thus making
very unlikely a reversed closure, while $\g(0)$ decreases. After a
straightforward calculation we get

\be \g(q)=1-{2\over \pi}\int_{\rm arcsin \sqrt q}^{\pi/2}
\exp \l(-{1\over 2}\l(h/\sin t\r)^2\r) dt. \ee
For large threshold the closing probability differs from 1 by a value
that cancels very fast, as $\exp\l(-h^2/2\r)$.
 
The attractors of the first type (such that $\G=R \G$) are completely
destroyed in this way, while attractors of the second type do not live in
pairs anymore, and the distribution of the weights is of the Random Map type.

\section{Numerical results}
\subsection{Distribution of the overlap and closing probabilities}
Our first aim was to compare the distribution of the overlap
predicted by the annealed approximation with the same distribution in the
quenched system. As we wrote, the analogy holds if we measure the
overlap between configurations only along the trajectories that are
not yet closed when we do the measurement. Under this condition, we
computed the distribution of the overlap $q(t,t+l)$ for $l$ fixed and
$t$ large enough to suppose that the distribution has attained stationarity.

The exponent $\a(q)$ of the distribution of the overlap is defined by
the equation $P_N(q)=C_N(q)\exp\l(-N\a(q)\r)$, where the factor
$C_N(q)$, proportional to $1/\sqrt N$, comes from the Stirling
expansion of the binomial coefficient. Thus we computed $\a(q)$ using
the formula

\be \a(q)=-{1\over N}\l(\log \l(P_N(q)\r) +{1\over 2}\log N\r). \ee
The logarithmic term must not be subtracted when $q$ is equal to 0 or 1,
because in this case the $1/\sqrt N$ factor is no more present in the
expansion of the binomial coefficient, and so we did not consider it
for $q=0$ and 1, interpolating linearly between the two formulas
for values of $q$ between 0 and $0.1$ and between $0.9$ and 1. In such
data analysis we neglect terms of order $1/N$ (there is also an
unknown coefficient in the expression of the probability $P(q)$), and
the agreement between the annealed prediction for $\a(q)$ and the
quenched data, compared in Figure 1, is then very satisfactory even for a
system of such a small size (we considered $N=20$). When $l$, the
temporal distance between configurations, is small, there are some
discrepancies (for instance, for $l=2$ the quenched distribution is
much broader 
than expected, and the closing probability is consequently much higher), but
when $l$ is large the agreement improves (in particular, the variance
of the distribution and the exponents $\a(0)$ and $\a(1)$ of the
closing probabilities coincide within the errors with the predicted
values). This fact sustains our interpretation that the annealed
approximation is valid when the temporal distance is large, so that
the system has lost memory of the details of its evolution
\cite{BP0,BP1}.

\begin{figure}
\centering
\epsfysize=12.0cm 
\epsfxsize=15.0cm 
\epsffile{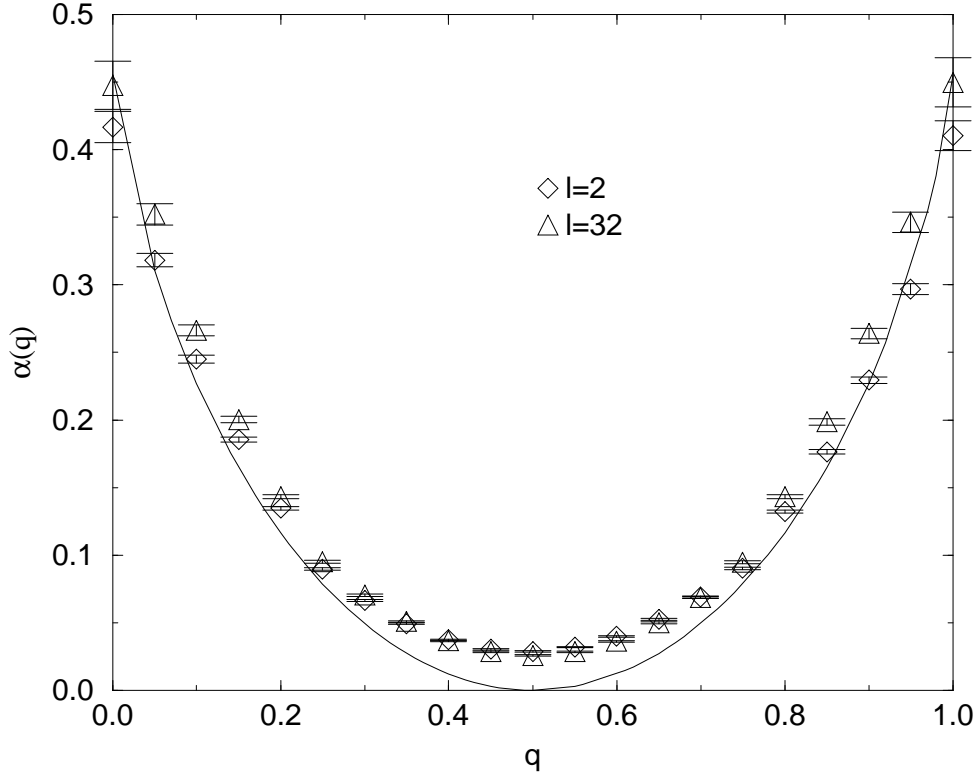}
\caption{\it Exponent $\a(q)$ computed from the equation $P_N(q)=C_N(q)
  \exp\l(-N\a(q)\r)$, where $P_N(q)$ is the stationary overlap
  distribution, with both $q=q(t,t+2)$ (diamonds) and $q=q(t,t+32)$
  (triangles). The solid line shows the annealed prediction. The
  agreement is better in the second case, when the temporal distance
  between configurations is larger. System size is $N=20$ and $h=0$. }
\label{fig1 }
\end{figure} 

Next we measured the closing probability $\pi_N(t,t+l)$.
Figure 2 represents this quantity as a function of $t$ for different values
of $l$, kept fixed. The statistic errors are large, but it appears
that $\pi(t,t+l)$ reaches a value approximately stationary in $t$, in
agreement with the annealed prediction, when $l$ is large (in figure
2b we have $l=11$) but when $l$ is small (in figure 2a
$l=2$) the closing probability reaches a maximum value and then
decreases, as a function of $t$. We already observed this
kind of non monotonic behavior of the closing probabilities in simulations
of Kauffman model. In both cases we interpret the decreasing part of $\pi(t)$
as due to the opening condition: the condition that the trajectory is
not closed up to time $t$ selects, as $t$ grows, trajectories which
are more and more
unlikely to close. The opening condition cannot be imposed in the annealed
scheme, because we consider the stochastic process $d(t,t+l)$ with $l$ fixed
and we cannot control $d(t,t')$ for generic $t$ and $t'$. So the
annealed scheme must be modified to take into account this fact
\cite{BP0}. But in asymmetric neural networks, differently from what we
observed in Kauffman networks, the opening condition seems to be
irrelevant when  the temporal distance $l$ is large, and the closing
probability appears to reach in this case an approximately stationary
value.

\begin{figure}
\centering
\epsfysize=12.0cm 
\epsfxsize=15.0cm 
\epsffile{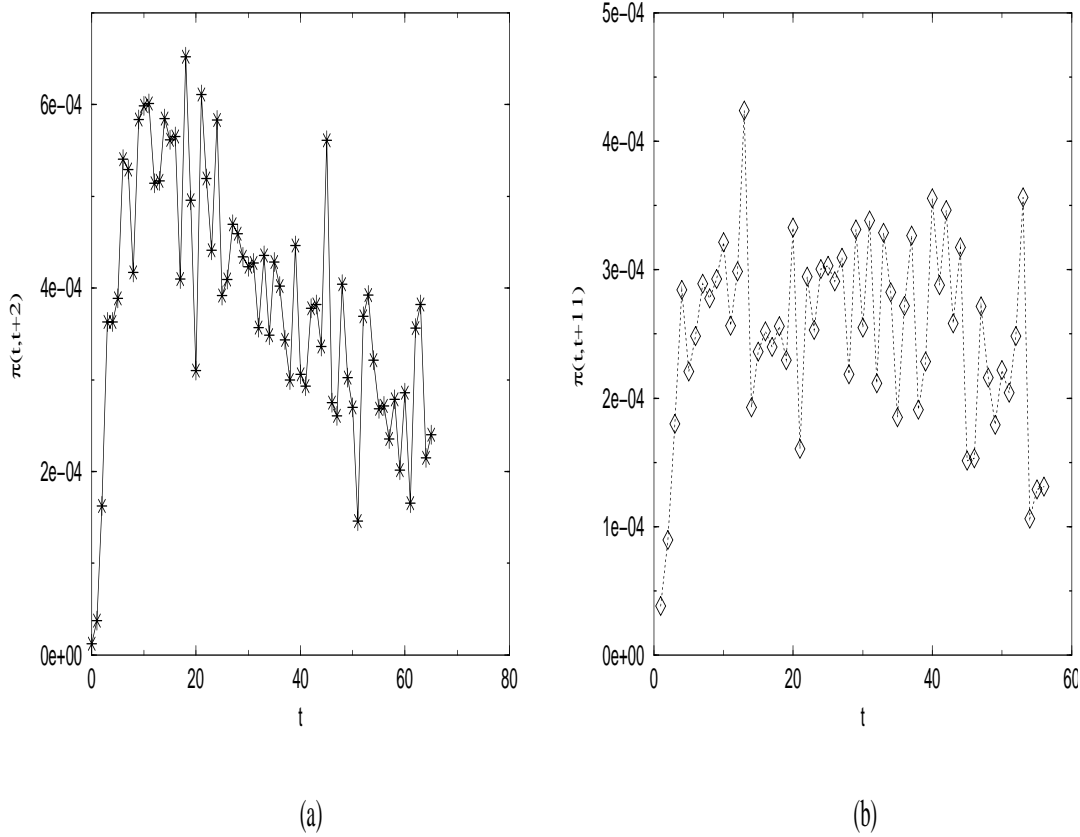}
\caption{\it Closing probability $\pi_N(t,t+l)$ as a function of $t$, for
  $l=2$ (a) and $l=11$ (b). System size is $N=20$ with $h=0$. }
\label{fig2 }
\end{figure} 

The non-stationarity of the distribution of $q(t,t+2)$ shows the
existence of memory effects in the model: the statistical properties
of $q(t,t+2)$ still depend on $t$, even after an arbitrarily long  transient
time. It would be interesting to find out whether the lack of time
translation invariance in the system with completely uncorrelated
couplings has some relation with aging in the relaxational dynamics of
the SK spin glass model \cite{CK}. In the present case, however,
the lack of time translation invariance is only a minor
effect and does not prevent the overlap $q(t,t+l)$ from reaching a
stationary distribution for $l$ large enough. The macroscopic
properties of the dynamics can be predicted, in good
agreement with numerical results, also neglecting this effect at all.

We conclude this subsection showing a plot of the integral
closing probability $\tilde\pi_N(t)$, defined as
\be \tilde\pi_N(t)=\sum_{t'=0}^{t-1} \pi_N(t',t)\ee
this is the probability that a trajectory not closed at time $t-1$ closes at
time $t$). In Kauffman networks, this quantity is non-monotonic as a
function of $t$: it increases to a maximum value and then decreases with $t$.
On the other hand, from the annealed approximation we would expect it to
increase linearly with $t$ in the stationary state. In asymmetric
neural networks we found that the integral closing probability
increases monotonically with $t$. After a transient phase of very
fast increase it slows down, and asymptotically it appears to behave
as a power law. For the largest systems that we simulated our data are
very noisy, and we could fit the asymptotic $t$ behavior only for
$N=20$, finding that the best fit exponent of the power law is
approximately $0.6$. So, at least for systems of not very large size,
deviations from the annealed approximation are present also in this
case.

\begin{figure}
\centering
\epsfysize=12.0cm 
\epsfxsize=15.0cm 
\epsffile{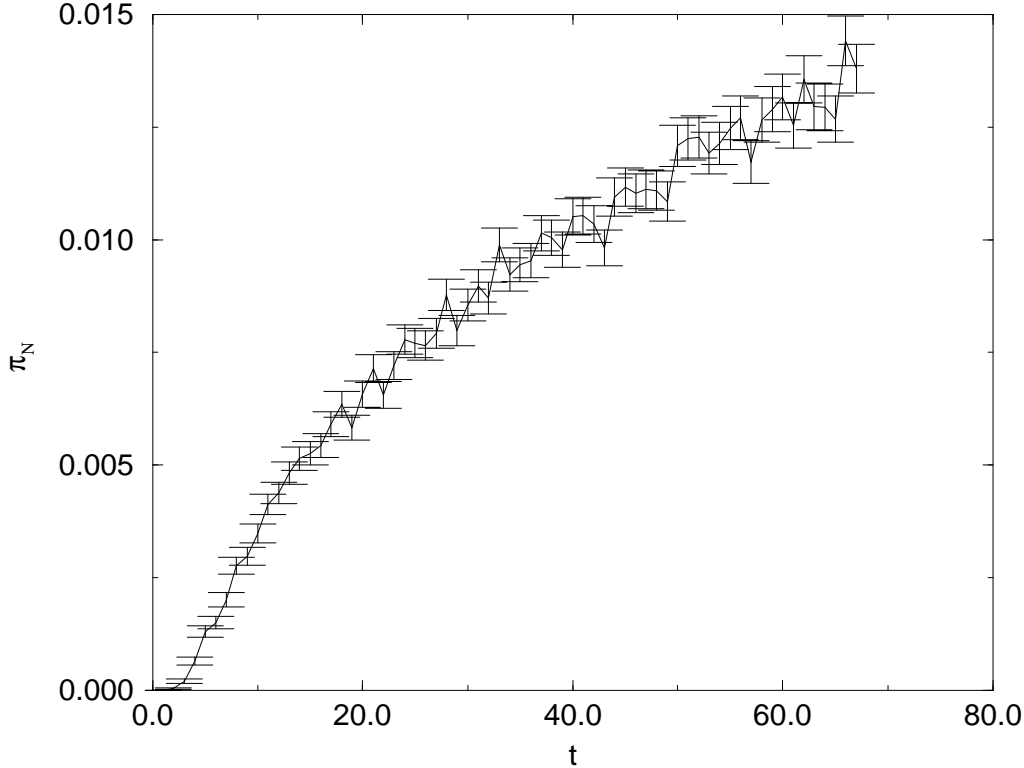}
\caption{\it Integral closing probability, $\pi_N(t)=\sum_{t'}\pi_N(t',t)$ as
  a function of $t$ in a system of size 20 with $h=0$ }
\label{fig3 }
\end{figure} 

\subsection{Properties of attractors: zero threshold}
To obtain the first three moments of the distribution of the attraction
basins we followed the method indicated in \cite{DF1}. For every value
of the parameters $N$ and $h=0$ we generated at random 2000 networks,
extracting the synaptic couplings with Gaussian distribution, and we
simulated four randomly chosen trajectories on each of them.

The average weight of the basins, $\la Y_2\ra$, was estimated from the
probability that two different trajectories end up on the same
periodic orbit. In general \cite{DF1},
$\la Y_n\ra$ can be measured as the probability that $n$ different initial
configurations evolve to the same attractor. Simulating four initial
configurations it is also possible to measure $\la Y_2^2\ra$ as the probability
that each of two pairs of configurations end up on a same attractor, the two
attractors being either different or equal.

Figure 4 shows data which report the behavior of the moments of
attraction basin distribution for systems of different size, $N$. It
can be seen that they rapidly converge to the predictions of the
annealed approximation, corrected to take into account the reversal symmetry.

\begin{figure}
\centering
\epsfysize=12.0cm 
\epsfxsize=15.0cm 
\epsffile{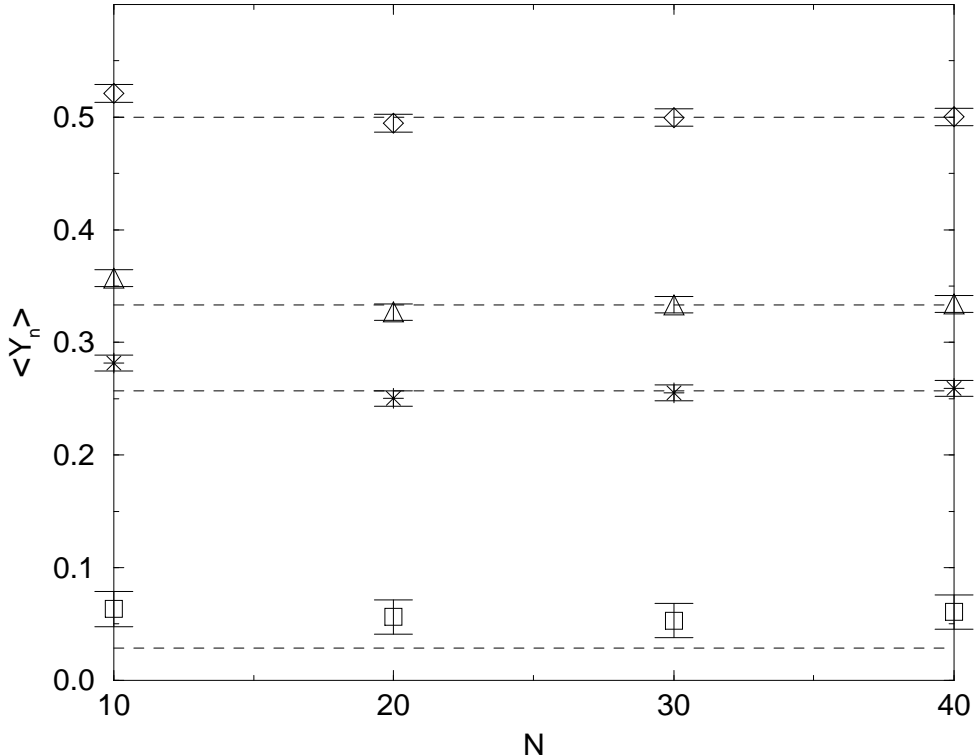}
\caption{\it Moments of the distribution of the attraction basin weights
  versus system size $N$ for $h=0$: $\la Y_2\ra$ (diamonds); $\la
  Y_2^2\ra$ (squares); $\la Y_3\ra$ (triangles)  and $\la Y_4\ra$
  (stars). The dotted lines show the predictions of the annealed
  approximation. }
\label{fig4 }
\end{figure}

The average cycles length increases exponentially with system
size, $\la L\ra\propto \exp(\a_L/2 N)$. The exponent $\a_L/2$ is less
than $\log 2/2=0.347$, as it would be in a completely random map. Its value
$\a_L/2=0.224$ is in good agreement with the prediction of the annealed
approximation, $\a/2=0.228$ (the small discrepancy could be a finite
size effect, as the exponent estimated from numerical data increases
when only the largest systems are considered). Figure 5 shows the
average length of the cycles versus system size.

\begin{figure}
\centering
\epsfysize=12.0cm 
\epsfxsize=15.0cm 
\epsffile{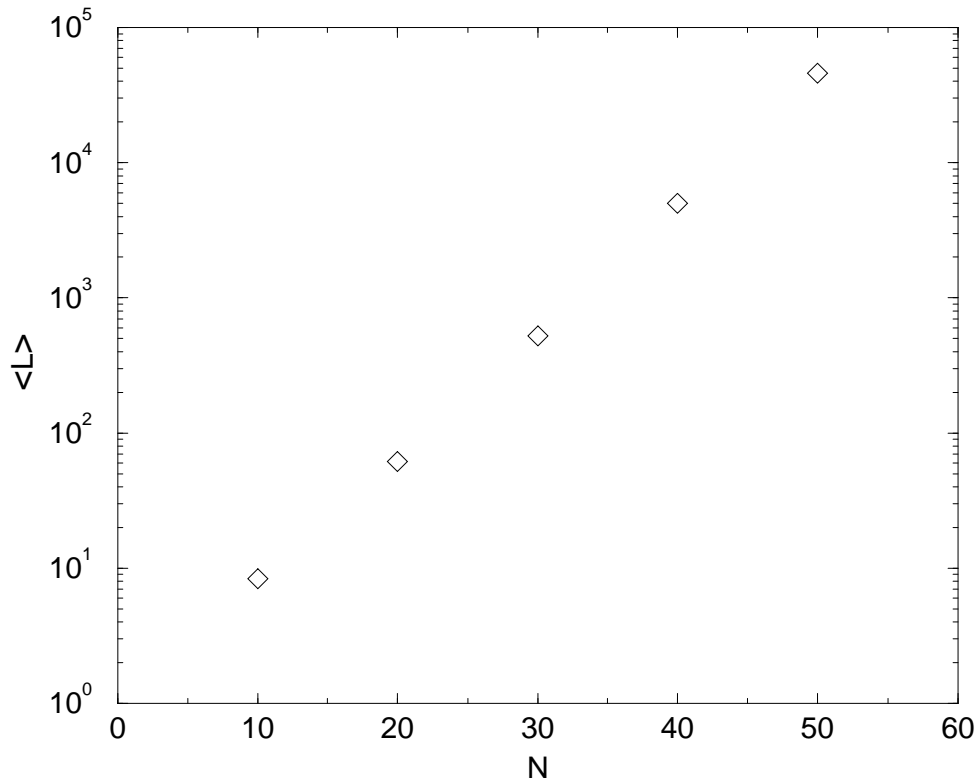}
\caption{\it  Average length of the cycles as a function of system size for $h=0$. }
\label{fig5 }
\end{figure} 

The distribution of cycle length is much broader than it is expected
on the basis of the annealed approximation, and asymptotically
behaves as a stretched exponential:

\be \Pr\l\{ L > l\r\}\approx \exp\l(-(l/\t_N)^{\g_N}\r).\label{distr}
\ee

As discussed in the previous sections, the distribution is different
for the two different types of cycles. We considered only odd
cycles, in order to select only attractors of the second type, and we checked
that the scale of the distribution, $\t_N$, increases exponentially
with $N$, $\t_N\propto \exp\l(\a_P /2 N\r)$, where the exponent
$\a_P$ coincides with $\a_L$ within the errors. On the other hand
the exponent $\g_N$ of the stretched exponential, for which the annealed
approximation predicts the value 2, is instead less than 1 for all of the
system sizes that we examined, but it appears to increase slightly as
$N$ grows (though are data about this point are very noisy),
so that it is possible that this discrepancy shall disappear in the
infinite size limit.

The fact that we find $\g_N$ less than 1  appears challenging also
because the distribution of the closing time ({\it i.e.} the sum of
the transient time plus the length  of the cycle), which should have the same
behavior of the distribution of cycle length, according to the
annealed approximation, is indeed much steeper: it can be fitted to a
stretched exponential of the same form (\ref{distr}), but with a much
larger exponent $\g'_N$. For instance, for $N=20$, we find
$\g'_N=1.9$, in good agreement with the annealed prediction, while the
value of $\g_N$ is $0.69$. 

\subsection{Properties of attractors: broken symmetry}
When we consider the evolution equation (\ref{evolh}) with a threshold $h$, the
reversal symmetry is explicitly broken and the distribution of the
weights is the same as in the usual Random Map.

Figure 6 shows the behavior of the first moments of the distribution
of the weights
as a function of system size $N$ for $h=0.1$. Such threshold is so
small that it modifies the value of the exponent $\a$ by less than 2
percent. The annealed approximation predicts in this case  $\a=0.448$, to be
compared to the value $0.455$ found with zero threshold. The
prediction is in good agreement with numerical simulations: a fit of
the average length of the cycles gives $\a_L=0.44$.
For such a small threshold we can observe traces of the broken symmetry
present as finite size effects: the moments of the distribution at small $N$
fall below the Random Map values, even if of a very small amount,
and then increase to those values, which are maintained asymptotically
in system size.

\begin{figure}
\centering
\epsfysize=12.0cm 
\epsfxsize=15.0cm 
\epsffile{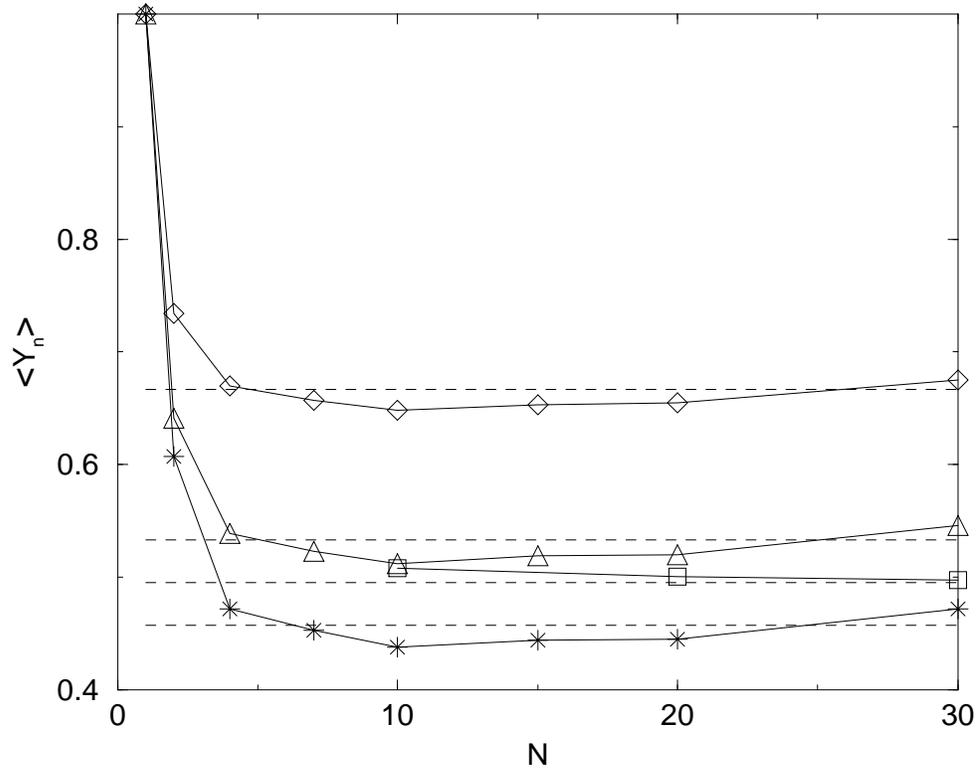}
\caption{\it Moments of the distribution of the attraction basin weights
  versus system size  $N$ for $h=0.1$. : $\la Y_2\ra$ (diamonds); $\la
  Y_2^2\ra$ (squares); $\la Y_3\ra$ (triangles)  and $\la Y_4\ra$
  (stars). The dotted lines show the predictions of the annealed
  approximation. }
\label{fig6 }
\end{figure} 

On the other hand, when the threshold is larger, we do not see at all the
signs of the symmetry on the distribution: for $h=1$, the average basin weight
decreases monotonically from the value 1 at small $N$ toward the Random Map
value $\la Y_2\ra=2/3$. For such a threshold the average length of the
cycles still behaves exponentially with $N$, but the exponent $\a$ is
very small
and power law corrections have important effects also for systems large
to simulate, as it appears from the fact that the best fit exponent
depends significantly on system size (it decrease as system size
increases), and we could not estimate it accurately. Nevertheless, the
agreement between the annealed approximation, which predicts
$\a=0.128$, and the numerical result $\a_L=0.15$, is worse than in the
previous cases but still not bad.

\section{Summary and conclusions}
In this work we used a stochastic scheme, based on the closing
probabilities and on their approximation by means of a Markovian
stochastic process, in order to compute the properties of the
attractors in fully asymmetric neural networks. The fundamental
hypothesis behind this approximation is that the system forgets fast
enough the details of its past evolution, so that a one step memory is
already enough to describe the gross features of the dynamics.
Our method is
able to predict very satisfactorily the $N$ behavior of the typical lengths
of the cycles and typical transient times, the number of cycles, the
distribution of their attraction basin weights and also the main
features of the distribution of the distances. On the other hand, the
approximation fails to predict the shape of the distribution of cycle length,
which is much broader than we would expect.

The average number of cycles had already been exactly computed, and perhaps
other quantities can be exactly computed in this model, but the
present method has the advantage of being very simple, and we hope that
it can be applied to more complex situations. In particular with this
method we argue that the distribution of attraction basins is, for
disordered dynamical systems that are ``chaotic enough", always equal to
the one computed by Derrida and Flyvbjerg for the case of an uniform
Random Map \cite{DF}.

A possible extension of our method, that we consider very interesting
and that we plan to pursue further, is towards the study of
neural networks with finite symmetry. Numerical studies suggest that
such systems undergo an abrupt change of dynamical regime when the
symmetry $\eta$ is changed \cite{Nu}, but a ``mean field" description
of this transition from an ordered behavior to chaos is still lacking.
The possibility that such a change can be characterized as a
transition between memory and loss of memory is very appealing.
Memory effects are more and more important for networks with non-zero
coupling symmetry (till the well-known aging properties of the SK
model are approached). Because of these effects, it is necessary to modify
our method also to study the chaotic regime of the model (low
symmetry). Technically this is not an easy task, since for non zero
symmetry correlations arise both between the local fields (see
equations (\ref{h+-})) of different neurons and, more difficult to
treat, between synaptic couplings and dynamical variables. The latter
introduce an effective interaction between the state of an element at
two different time steps $t$ and $t+2$, so that in the dynamics  also
an effective gradient flow is present, and,
if the annealed approximation can describe this situation, it will be
necessary to take into account also this information, aside the crude
distance, to make the annealed scheme useful.

Studying this family of models it is also possible, varying the
continuous parameters $\eta$ and $h$, to go from the
distribution of the attraction basins typical of the Random Map to the one
typical of Spin Glasses, thus the study of the general model would
shed some light on the relation between the two kinds of distributions.

Memory effects are probably responsible of the discrepancy between the
prediction of the annealed approximation and the observed distribution
of cycle length. In fact,  the distance $d(t,t')$ does not reach a
stationary distribution, if we impose the condition that the
trajectory is not yet closed before the measure. We think that this
condition, which can not be imposed in our computation, selects
trajectories that are less and less likely to close. As a result, the integral
closing probability, which is our main tool in the computation,
increases as a function of $t$ slower than expected.
It is possible that this effect shows up only at small $l$ and disappears
in the infinite size limit (an hint of this could be the fact
that the distribution of cycle length decays faster in this
limit), but it is also possible that, as in the case of Kauffman model
that we previously studied, some corrections to the annealed picture
are necessary also in the infinite size limit. However, we think that
these results show that
the annealed approximation is an useful tool to investigate in a simple
way the properties of attractors in disordered dynamical systems. 

\section*{Acknowledgments}
We are indebted to Angelo Vulpiani for addressing us to this model. U.B. is
pleased to thank Peter Grassberger and Heiko Rieger for interesting
discussions.


\end{document}